\documentclass[intlimits,twoside,a4paper]{article}

\usepackage[cp1251]{inputenc}

\usepackage[eqsecnum]{cmpj3}

\usepackage{bm}

\issue{2017}{20}{3}{33604}
\doinumber{10.5488/CMP.20.33604}

\title[Cylindrical nanopore-colloid interface]%
{Model charged cylindrical nanopore in a colloidal dispersion: charge reversal, overcharging and double overcharging}

\author[E. Gonz\'alez-Tovar, M. Lozada-Cassou]{E. Gonz\'alez-Tovar\refaddr{label1}, M. Lozada-Cassou\refaddr{label2}\thanks{Corresponding author}}
\addresses{
\addr{label1} Instituto de F\'\i sica, Universidad Aut\'onoma de San Luis Potos\'\i , 78000 San Luis Potos\'\i, S.L.P., M\'exico
\addr{label2}Instituto de Energ\'\i as Renovables, Universidad Nacional Aut\'onoma de M\'exico (UNAM), 62580 Temixco, Morelos, M\'exico
}
\date{Received July 29, 2017, in final form August 7, 2017}

\begin{document}

\maketitle

\begin{abstract}
Using the hypernetted-chain/mean spherical approximation (HNC/MSA) integral equations we study the electrical double layer inside and outside a model charged cylindrical vesicle (nanopore) immersed into a primitive model macroions solution, so that the macroions are only present outside the nanopore, i.e., the vesicle wall is impermeable only to the external macroions. We calculate the ionic and local linear charge density profiles inside and outside the vesicle, and find that the correlation between the inside and outside ionic distributions causes the phenomena of overcharging (also referred to as surface charge amplification) and/or charge reversal.  This is the first time overcharging is predicted in an electrical double layer of cylindrical geometry. We also report the new phenomenon of double overcharging. The present results can be of consequence for relevant systems in physical-chemistry, energy storage and biology, e.g., nanofilters, capacitors and cell membranes. 
\pacs 68.08.Bc, 87.15kt, 82.47.Aa
\end{abstract}

This paper is devoted to the memory of \textbf{Professor Jean-Pierre Badiali}, with whom, one of us (ML-C), expended many hours of scientific discussions, at the \textit{Universit{\'e} Pierre et Marie Curie}, in Paris and in my office, in Mexico City, and personal and philosophy discussions at  \textit{Les Deux Magots}.

\section{Introduction}

The ionic distribution of an electrolyte next to a charged colloid or surface gives rise to the well-known concept of electrical double layer (EDL). The study of the EDL is a classic subject in physical-chemistry due to its ubiquity and undoubtful relevance for many fundamental problems and uncountable applications in colloid science and technology~\cite{hunter,wennerstrom,muller92,kovac92}. Most of the equilibrium and transport behavior of colloidal matter, e.g., stability and electrokinetics, is governed by the electrolytic structure around the surface of a macroparticle, and, accordingly, a great deal of attention has been paid to obtain a reliable description of the EDL~\cite{Dukhinbook, delgado2005}. As a result, and departing from the seminal work by Helmholtz, Gouy, Chapman, and many others~\cite{verwey48}, the theory of this physico-chemical charge distribution has evolved substantially up to its present state, in which an adequate knowledge of the associated phenomenology has been established. However, and in spite of the progress, two new research venues have revivified this area in the recent times. In the first place, a new and intriguing phenomenon has been reported quite recently, namely  overcharging (OC)~\cite{jimenez04}. Other conspicuous interface phenomena observed lately are charge reversal (CR)~\cite{jimenez04} and charge inversion (CI)~\cite{jimenez04}. CR occurs when the excessive presence of counterions close to an electrified interface results in an effective, or accumulated, charge (of the interface plus the adsorbed ions) with a contrary sign to that of the bare charge of the macroparticle/surface. This reversal of the electrical field, in turn, produces an additional charge inversion, but this time the accumulated charge is of the same sign as in the macroparticle/surface. On the other hand, the term OC applies when the unexpected adsorption of coions \textit{increases the magnitude} of the accumulated charge above that of the native one (without an inversion of sign). In the immediate past, significant interest regarding this last pair of charge anomalies has arisen in the literature since an understanding of the phenomena of CR and OC could shed light on new and fascinating physical mechanisms acting in coulombic fluids~\cite{vatamanu,Messina2007,chialvo2011,olvera2014,podgornik2014,wang2016,seanea2017,seanea2017b}.

It must be noted that, in the past, OC was also identified in the literature with the name of surface charge amplification (SCA)~\cite{chialvo2008}, to avoid confusion when the same term is used as synonymous of CR or CI. The preposition ``over'' refers to ``above or higher than something else'', thus ``overcharging'' means electrical charge above the original charge. Nevertheless, in some literature the word overcharging has been indistinctly used as equivalent to ``charge reversal'' and/or ``charge inversion'', e.g., ``\textit{The terms ``overcharging'', ``charge reversal'', and ``charge inversion'' are more or less synonymous\,\ldots}'' \cite{lyklema2006}. In electrical double layer phenomena, these three terms have clear distinct significances, as we have defined them above. Thus, in this work overcharging means a local charge of a higher magnitude than that of the native one. CR was investigated previously~\cite{vlachy1982,patra1999,bhuiyan2005,patra2005,outhwaite2013}, although frequently as an implication of the inversion of the potential of mean force (i.e., therein no induced charge was calculated), and sometimes it is incorrectly referred to as overcharging.

A second theme of interest in the EDL area is the consideration of more realistic representations of double layer systems, i.e., with sophisticated geometries and/or with additional features or interactions, for example, with dielectric contrast, dispersion forces, etc. Along the same line, in the last years, several papers have been published dealing with EDLs of non-simple forms~\cite{vlachy1986,attard96,Yu1, seanea2017, seanea2017b} and/or that explore the effect of a molecular solvent, non-spherical ions, van der Waals and image forces, and discrete or random surface charges on the EDL structure~\cite{patra2009,olvera2014, podgornik2014, wang2016}. Interestingly, some of these last reports have delved into the influence of such interactions on the fate of CR and OC~\cite{olvera2014, podgornik2014, wang2016}.

Various reviews have also pointed out a discernible link between the electrical double layer and biological matter~\cite{bachbook}. Thus, any advancement in the comprehension of more sophisticated EDL systems could have a great impact on the study of many biomolecular materials and phenomena, e.g., cellular membranes, blood cells, key-lock mechanisms, viral encapsidation and much more~\cite{gelbart2000, teubl}. In this scenario, it would be desirable to extend the use of the traditional experimental, computer simulation and theoretical techniques to analyze the behavior of more complex EDL systems under conditions of overcharging and/or charge reversal. A good example of such could be the examination of ionic liquids confined in a spherical shell or inside cylindrical or slit pores.

With this idea in mind, in this work we undertake a theoretical investigation of the EDL  inside and outside a model cylindrical nanopore or vesicle of cylindrical shape, immersed into a primitive model (PM) colloidal dispersion or macroions solution, so that the nanopore wall is \textit{impermeable only to the external colloidal particles}, i.e., they are present only outside the nanopore, with special emphasis on the behavior of the induced charge within the surrounding ionic atmosphere. This is an EDL problem that has been scarcely pondered in the literature~\cite{Yeomans1,aguilar2007} and with a plausible relevance in physical-chemistry, biology and materials science. 

This paper is organized as follows. First, we present the model of the nanopore-electrolyte system to be used, the hypernetted-chain/mean spherical approximation (HNC/MSA) theoretical description of the associated EDL and its numerical method of solution. Also, in the same section, we introduce the concept of accumulated charge density, which plays a key role in our study of overcharging and charge reversal. In the rest of the article we proceed to describe and discuss our results for the EDL structure around the model vesicle, i.e., the ionic and local linear charge density profiles inside and outside the cylindrical vesicle, with a particular focus on OC and CR. We then finish with our conclusions and closing remarks. We stress that while we do not present computer simulation results for our system, the HNC/MSA theory has been extensively and successfully compared with density functional theories, Monte Carlo and/or molecular dynamics simulations for a large variety of charged homogeneous and inhomogeneous fluids~\cite{Degreve1,Degreve3,Deserno1,manzanilla2011b,levin2016,seanea2017b,seanea2017}. Hence, we esteem that the validity and predictions of the HNC/MSA integral equations formalism applied to homogeneous or inhomogeneous PM systems has been sufficiently demonstrated.
\newpage
\section{Model and theory} 

\begin{figure}[!b]
\vspace{-5mm}
\begin{minipage}{0.9\textwidth}
     \centering
\includegraphics[width=.7\textwidth]{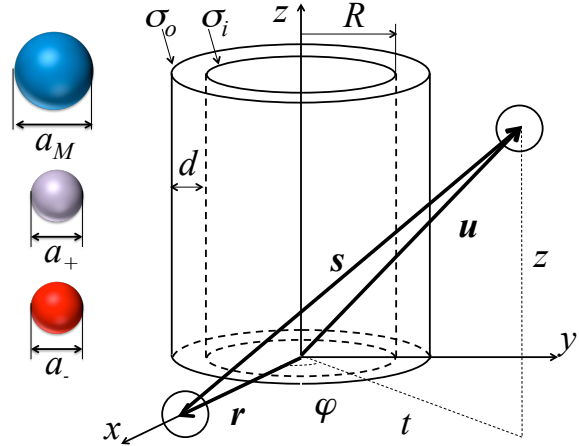}
\end{minipage}
\vspace{-5mm}
\caption{\label{Fig1} (Color online) Model scheme of a colloidal primitive model (CPM) next to a charged cylindrical vesicle (pore). Colloidal particles in an electrolyte solution, next to a cylindrical vesicle, of internal radius~$R$ and thickness $d$. The colloidal particles are at a finite volume fraction,  $\phi\equiv \frac{1}{6}\piup\rho_{\text M}a_{\text M}^3=0.01$, with bulk concentration $\rho_{\text M} =0.0045$~M, diameter $a_{\text M}=4.5a$, and $\sigma_{\text o}$ and $\sigma_{\text i}$ being the outside and the inside surface charge densities of the vesicle (pore) walls. The salt is a 1:1, $0.05$~M RPM electrolyte, with $a_{-}=a_{+}\equiv a=4.25$~{\AA}. The big cyan spheres are negatively charged colloidal particles and the little red and purple spheres are anions and cations, respectively. }
\end{figure}

Our model for a cylindrical vesicle (or cylindrical nanopore) immersed into a colloidal suspension consists of a charged and hard cylindrical pore bathed by a three-component coulombic fluid mimicked via the colloidal primitive model (CPM)~\cite{manzanilla2011a}, that is, hard sphere ions with point charges at their centers supported by a continuous solvent (see figure~\ref{Fig1}). More explicitly, the pore is infinitely long and has an internal radius $R$ and thickness $d$, and is bearing internal and external surface charge densities $\sigma_{\text i}$ and $\sigma_{\text o}$, respectively. Here, we have restricted ourselves to the case $\sigma_{\text i}=\sigma_{\text o}$. Three ionic species compose the macroions solution or colloidal suspension. They have diameters and charges $a_j$ and $q_j=ez_j$, respectively, where $e$ is the protonic charge and $z_j$ is the valence of the species $j$, such that $j=1,2,3$. In this investigation we will consider two small and equal-size ionic species with symmetric charges, i.e., $a_1 = a_2 = a$ and $z_1=-z_2$, plus a third species of macroions, such that $a_3> a$. For completeness, we take $z_2 z_3 >0$, and, when the pore is charged, the macroions and the ions of species $j=2$ are its counterions, i.e., $\sigma_{\text i}z_3<0$. All the system has a dielectric constant $\epsilon$ and is at a temperature $T$. Accordingly, the interaction potential between the electrolytic ions is 
\begin{align}
	U_{mj}(r)=
	\begin{cases}
		\infty, & \text{for} \quad r < (a_m+a_j)/2,  \\
		 \dfrac{z_m z_je^2}{\epsilon r}\,, & \text{for} \quad r \geqslant (a_m+a_j)/2,
	\end{cases}
\end{align}
with $m,j=1,2,3$; whereas, the cylinder-ion potential is 
\begin{align}
	U_{cj}(r)=
	\begin{cases}
		-\frac{4\piup e}{\epsilon}z_j \left[ R \sigma_{\text i} \ln R + (R+d) \sigma_{\text o} \ln (R+d) \right]\\
		+ \frac{4\piup e}{\epsilon}z_j \left[ R \sigma_{\text{i}} + (R+d) \sigma_{\text{o}} \right] \ln \infty, & \text{for}\quad r \leqslant R-(a_j/2),  \\		
		\infty, & \text{for}\quad R-(a_j/2) < r < R+d +(a_j/2),  \\
		-\frac{4\piup e}{\epsilon}z_j \left[ R \sigma_{\text{i}} + (R+d) \sigma_{\text{o}} \right] \ln r \\ 
+ \frac{4\piup e}{\epsilon}z_j \left[ R \sigma_{\text{i}} + (R+d) \sigma_{\text{o}} \right] \ln \infty, & \text{for}\quad r \geqslant R+d+(a_j/2).
	\end{cases}
\end{align}

The establishment of the HNC/MSA description of the electrical double layer surrounding a cylindrical pore proceeds from the homogeneous and multicomponent Ornstein-Zernike equation (for a mixture with four constituents) 
\begin{align}
	h_{\alpha j}({r}_{12})= c_{\alpha j}({r}_{12})+ \sum_{m=1}^{4} \rho_m \int h_{\alpha m}({r}_{13}) c_{m j}({r}_{32})\,\rd V_3 \,,
\end{align}
where $\alpha ,j = 1,2,3,4$, and $h_{\alpha j}({r}_{12})$ and $c_{\alpha j}({r}_{12})$ are, respectively, the total and direct correlation functions for particles $1$ (species $\alpha$) and $2$ (species $j$), which are at a distance $r_{12}=\left| \vec{r}_2 - \vec{r}_1\right|$. Something equivalent applies for $h_{\alpha m}({r}_{13})$ and $c_{mj}({r}_{32})$. Besides, $\rho_m$ is the bulk number density of species $m$, and $\rd V_3$ is the volume element for particle $3$ (species $m$). In our system, species $1$, $2$ and $3$ are reserved for the macroions solution or colloidal dispersion components and species $4$ corresponds to cylindrical pores. Applying the direct method proposed by Lozada-Cassou~\cite{LozadaJCP81,LozadaJPC83} to derive a set of integral equations for the ionic distributions around a {\it single} pore located at the origin of coordinates, we take the limit $\rho_4 \to 0$ in the Ornstein-Zernike equation to get
\begin{align}
	h_{cj}(r)= c_{cj}(r)+ \sum_{m=1}^{3} \rho_m \int h_{cm}(u) c_{mj}(s)\,\rd V.
\end{align}
Note that, for simplicity, in the previous equation we have replaced the index $4$ by $c$ (cylindrical pore), redefined $r=r_{12}$, $u=r_{13}$ and $s=r_{32}$, and dropped the index 3 in $\rd V$. As it can be deduced from figure~\ref{Fig1}, $\rd V=t \rd t \rd\varphi \rd z$.

Employing the hypernetted-chain (HNC) closure for the pore-ion direct correlations, 
\begin{align}
	c_{cj}(r)= -\beta U_{cj}(r)+h_{cj}(r)-\ln [1+h_{cj}(r) ], 
\end{align}
and the mean spherical approximation (MSA) closure for a bulk electrolyte~\cite{kazuo77} to deal with the ion-ion direct correlation functions, namely
\begin{align}
	c_{mj}(s)=c_{mj}^{hs}(s)+c_{mj}^{sr}(s)-z_m z_j \frac{\beta e^2 }{\epsilon s} 
\end{align}
for $s \geqslant 0$, with $\beta = 1/(k_{\text B} T)$ and $k_{\text B}$ being the Boltzmann constant, we can write  
\begin{align}
	g_{cj}(r)&=\exp \left\lbrace -\beta U_{cj}(r) + \sum_{m=1}^{3} \rho_m  \int   h_{c m}(u) \,  c_{mj}^{hs}(s)\rd V   + \sum_{m=1}^{3} \rho_m  \int   h_{c m}(u) \,  c_{mj}^{sr}(s) \rd V  \right.  \nonumber\\ &\left.\quad\quad\quad+ \frac{2\piup\beta e^2 z_j}{\epsilon} \int_{0}^{\infty} \ln \left( \frac{r^2 + t^2 + | r^2 - t^2 | }{2}  \right)  \left[ \sum_{m=1}^{3} \rho_m z_m h_{c m}(t) \right]  t \rd t \right\rbrace.
\end{align} 
In the prior equation, $g_{cj}(r)=h_{cj}(r)+1$ is the well-known radial distribution function. Inserting the explicit MSA forms of $c_{mj}^{hs}(s)$ and $c_{mj}^{sr}(s)$ given in \cite{kazuo77}, we finally obtain 
\begin{align}
	g_{cj}(r)&=\exp \left\lbrace -\beta U_{cj}(r) + \sum_{m=1}^{3}  \rho_m \int_{t'}^{t''}\left[K_{mj}(r,t)+L_{mj}(r,t) \right] h_{cm}(t) t  \rd t\right.\nonumber\\&\qquad\quad\left.+ \frac{2\piup\beta e^2 z_j}{\epsilon} \int_{0}^{\infty} \ln \left( \frac{r^2 + t^2 + | r^2 - t^2 | }{2}  \right)  \left[ \sum_{m=1}^{3} \rho_m z_m h_{c m}(t) \right]  t \rd t  +A_j(r) \right\rbrace,
	\label{all1} 
\end{align}
where $t' = 0$ and $t'' = R-(a_j/2)$, for $0 \leqslant r \leqslant R-(a_j/2)$; or, else, $t' = R+d+(a_j/2)$ and $t'' = \infty$, for $\ r \geqslant R+d+(a_j/2)$. Above, $K_{mj}(r,t)$, $L_{mj}(r,t)$ and $A_j(r)$ are lengthy expressions resulting from the integration in the $z$ and $\varphi$ cylindrical coordinates. These three kernels are extensions, for size-asymmetric electrolytes, of the corresponding functions for equally-sized ions that can be consulted in \cite{Yeomans1,aguilar2007}. 

Equation~\eqref{all1} constitutes a complete set of non-linear integral equations for the ionic distributions $g_{c1}(r)$, $g_{c2}(r)$ and $g_{c3}(r)$, which has been solved numerically by means of a fast and efficient finite element method. In the recent past, this technique was advantageously used by the authors in several investigations of the electrical double layer in diverse geometries~\cite{Lozada-Cassou92a,manzanilla2013}. Once the ionic profiles have been determined, straightforward integrals of the $g_{cj}(r)$ provide important properties of the double layer. Particularly, we are interested in the {\it accumulated} surface and linear charge density profiles, $\sigma (r)$ and $\lambda (r)$, respectively. In physical terms, the function $\sigma (r)$ measures the charge per unit area contained in a cylindrical volume of radius $r$. Relatedly, $\lambda (r)$ quantifies that enclosed charge per unit length. This $\lambda(r)$ can be calculated by using 
\begin{align}
	\lambda (r)=2\piup e  \int_{0}^{r} \sum_{m=1}^{3} \rho_m z_m g_{cm}(t)  t\rd t.
\end{align}
It is easy to show that the quantities $\sigma (r)$ and $\lambda (r)$ are related by the expression $\sigma (r) = \lambda (r)/(2\piup r)$. For instance, here $\lambda_i = 2\piup R \sigma_{\text{i}}$ and $\lambda_{\text{o}} = 2\piup (R+d) \sigma_{\text{o}}$.

These two charge densities are fundamental in the present investigation given that they allow us to directly identify the occurrence of CR and/or OC.
However, for a pore-electrolyte solution, $\lambda (r)$ is a more appropriate form to describe OC and CR since in this system there is only one associated coordinate (the $z$ coordinate) that varies up to infinity. For this same reason, $\sigma (r)$ and $Q(r)$ (i.e., the enclosed charge up to a radial distance $r$) should be better choices to characterize the mentioned charge anomalies in planar and spherical geometries, respectively. In this way, in our cylindrical instance, for the region $0\leqslant r \leqslant R-(a/2)$, the condition $\lambda_{\text i} \lambda (r) < 0$ implies charge reversal at some point inside the pore, whereas the simultaneous observation of $\lambda_{\text i} \lambda (r) > 0$ and $\left| \lambda (r)\right| > \left| \lambda_{\text i} \right|  $ denotes overcharging somewhere inside; and, for $r \geqslant R+d+(a/2)$, $\lambda_{\text o} \lambda (r) < 0$ is equivalent to charge reversal in the outside region, and the pair of conditions $\lambda_{\text o} \lambda (r) > 0$ and $\left| \lambda (r)\right| > \left| \lambda_{\text o} \right|$ indicate overcharging outside the pore.


\section{Results and discussion}
We solve equations~\eqref{all1} for several values of $R$ and $\sigma_{\text i}=\sigma_{\text o}\equiv\sigma_{0}\in[0~\text{C/m}^{2},\,0.15~\text{C/m}^{2}]$ of the cylindrical vesicle, and two values of the macroions surface charge density, i.e., for $\sigma_{\text M}=-0.0558$~C/m$^{2}$ and $\sigma_{\text M}=-0.1534$~C/m$^{2}$, which correspond to $z_{\text M}=-4$ and $z_{\text M}=-11$, respectively. The thickness of the vesicle membrane is always taken to be $d=a$. The little anions and cations diameters were both taken to be equal to $a=4.25$~{\AA}, and the macroions diameter was taken to be $a_{\text M}=4.5a$. The macroions volume fraction was taken to be $\phi\equiv \frac{1}{6}\piup\rho_{\text M}a_{\text M}^3=0.01$, which corresponds to a macroions bulk molar concentration of $\rho_{\text M}=0.00455$~M. The salt is a 1:1, $0.05$~M electrolyte. The counterions species of macroions was taken to be of the same species as that of the salt cations. Hence, while the anions concentration is $\rho_-=0.05$~M, in all our results presented here, the cations concentrations are $\rho_+=0.06818$~M and $\rho_+=0.01$~M, for the $z_{\text M}=-4$ and $z_{\text M}=-11$ cases, respectively. In all the data reported here we have restricted ourselves to the case in which the vesicle membrane is impermeable to the macroions species, present only on the outside of the vesicle. Hence, we study the structure of the electrolyte inside the pore and that of the colloidal dispersion outside the vesicle. These two structures are correlated through the membrane of the pore, as one could expect since equation~\eqref{all1} is a set of non-linear integral equations, in which the inside and the outside of the pore fluid structures are interconnected; this has been shown to be very relevant in the past in confined systems~\cite{Yu2,Yu3,Degreve3}. In any case, notice that the HNC/MSA formalism of integral equations ensures the same chemical potential for the inside and the outside of the pore, and it has been shown to be in agreement with Monte Carlo and molecular dynamics simulations~\cite{odriozola2009,Yu1}. The dielectric constant of the solvent, taken to be as a uniform media, is $\epsilon=78.5$. To avoid image charges we have also chosen a dielectric constant of the pore walls and particles to be equal to that of the solvent. The temperature is $T=298$~K.

\begin{figure}[!t]
\vspace{-5mm}
 \begin{minipage}{0.49\textwidth}
     \centering
\includegraphics[width=.99\textwidth]{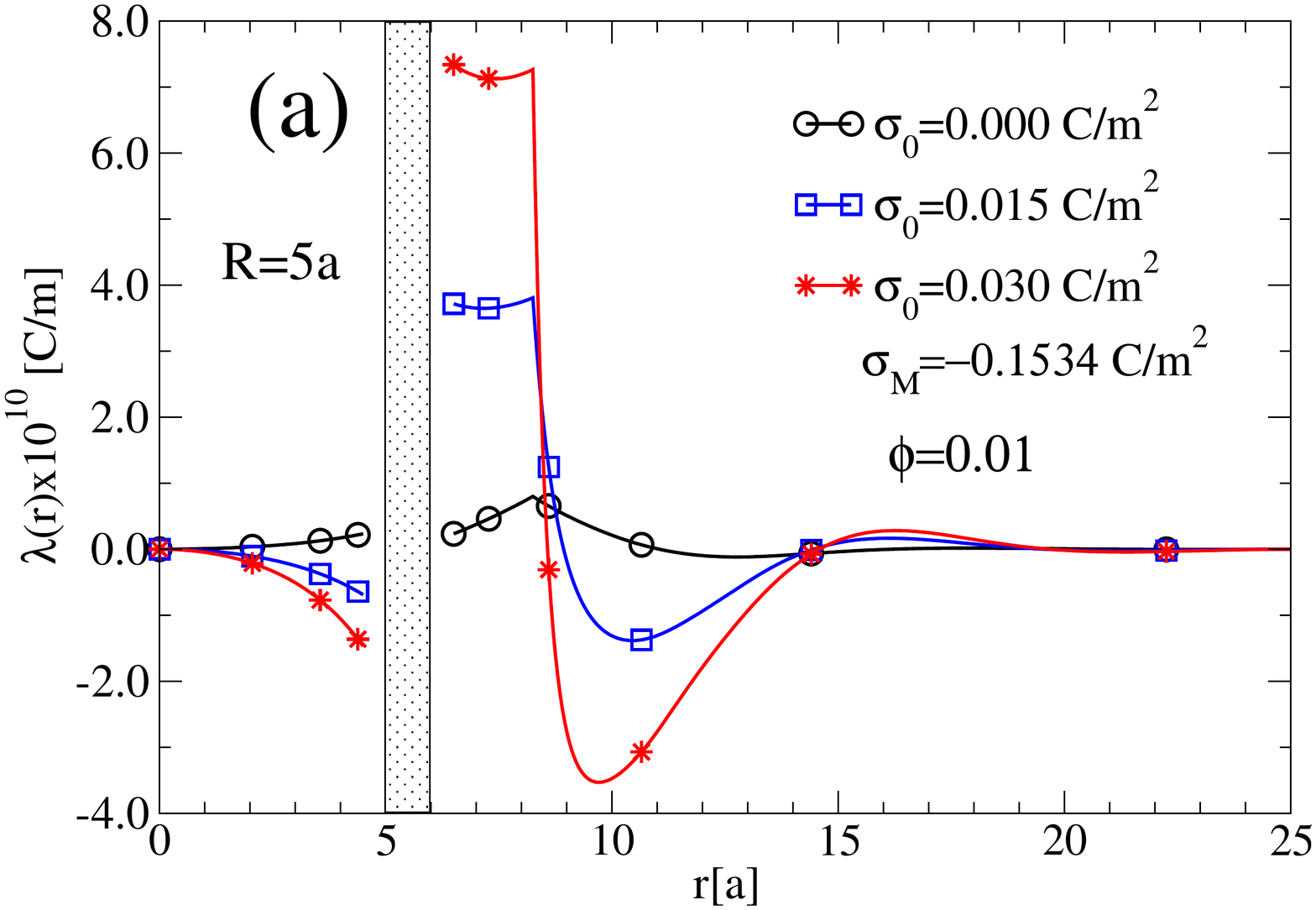}
\end{minipage}
   \begin{minipage}{0.49\textwidth}
   \centering
\includegraphics[width=.99\textwidth]{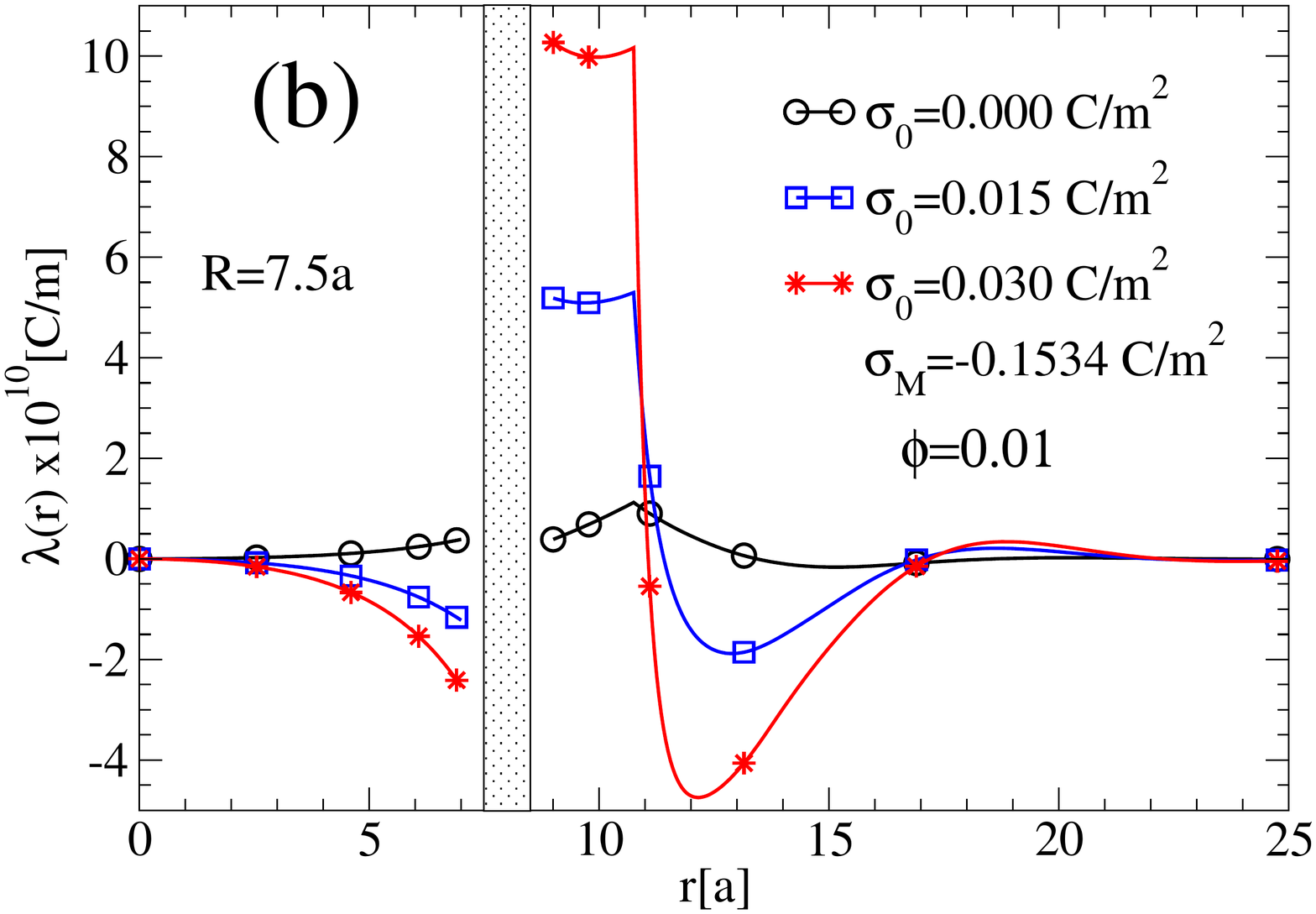}
     \end{minipage}
     \caption{\label{Fig2} (Color online) Induced linear charge density in a cylindrical vesicle (pore), as a function of the distance to the central azimuthal axis, for two values of the inner pore radius, $R$. The dotted bar represents the pore walls, of thickness $d=a$. To the left of this bar there are the induced linear charges densities inside the pore, and to the right those on the outside. The fluid is a colloidal particles dispersion in an electrolyte solution. The colloidal particles are at a finite volume fraction,  $\phi=0.01$ ($a_{\text M}=4.5a$, $\rho_{\text M}=0.00455$~M). The curves are for different values of surface charge densities on the pore walls, such that $\sigma_{0}=\sigma_{\text i}$. The salt is a 1:1, 0.05~M RPM electrolyte, with $a_{-}=a_{+}\equiv a=4.25$~{\AA}. }
\end{figure}

In figure~\ref{Fig2} we present our results for the induced linear charge density, $\lambda(r)$, as a function of the distance to the axial center of the nanopore, for two values of the inside pore radius, $R=5a$ and $R=7.5a$, $\sigma_{\text M}=-0.1534$~C/m$^2$ ($z_{\text M}=-11$), and three values of the membrane surface charge density, $\sigma_{\text{i}}=\sigma_{\text{o}}=\sigma_0=0.000$~C/m$^2$, 0.015~C/m$^2$, and 0.030~C/m$^2$. Notice that the total charge density on the membrane is $2\sigma_0$. For both values of the nanopore radius, and $\sigma_0>0$, the inside induced charge is negative. Hence, decreasing the net electrical field coming out of the nanopore, e.g., in figure~\ref{Fig2}~(a), \textit{without the electrolyte inside}, $\lambda(r=R+d)=4.8\times10^{-10}$~C/m, for the case of $\sigma_0=0.015$~C/m$^2$, and $\lambda(r=R+d)=9.6\times10^{-10}$~C/m, for the case of $\sigma_0=0.030$~C/m$^2$. However, the net induced charge in this point is lower, as seen in figure~\ref{Fig2}~(a). For the $R=7.5a$, shown in figure~\ref{Fig2}~(b), the situation is analogous. In figure~\ref{Fig2}~(b), for the $\sigma_0=0.000$~C/m$^2$ case we observe the opposite, the inside induced charge is positive and remains positive up to $\sigma_0\lesssim0.004$~C/m$^2$ (not shown). The same event is observed in figure~\ref{Fig2}~(a), for $R=5a$, since in this instance the inside induced charge is also positive and remains positive also up to $\sigma_0\lesssim0.004$~C/m$^2$, implying that this is a correlation effect between the outside and the inside ionic liquids, and not a confinement consequence. We will come back to this point below. 

Outside the pore, the induced linear charge density is oscillatory, i.e., is positive next to the nanopore, then negative (CR), and then positive again (CI). A point that we wish to emphasize is that, for $0$~C/m$^2$$\leqslant\sigma_0$ $\lesssim0.015$~C/m$^2$, we have $\lambda(r=R+d+a/2)\leqslant\lambda(r=R+d+a_{\text M}/2)$, i.e., there is an \textit{overcharging} of the nanopore wall. For higher values of $\sigma_0$, the overcharging disappears. The overcharging is generated by the cations brought next to the nanopore wall by the macroions, which in turn are attracted to the nanopore because they have an opposite charge to that of the nanopore, including the negative induced charge density inside the nanopore. There is also an entropy contribution due to the macroions size, i.e., by adsorbing on the nanopore wall, the system available volume is increased. The upward concavity of the induced charge density, outside the nanopore, and next to its surface, is due to the cations structure around the macroions, next to the nanopore wall. For $z_{\text M}=-4$ (not shown), we calculated the induced linear charge density for $R=5a$, and we found that this quantity is lower than that in figure~\ref{Fig2}~(a). The overcharging and hump observed in figure~\ref{Fig2}~(a) disappear, with the exception of $\sigma_0\approx0$~C/m$^2$.
It is interesting to note that while the linear charge density on the nanopore's external wall, $\lambda(R+d)\equiv\lambda_0$, is higher for the $R=7.5a$ than that for $R=5a$ (see figure~\ref{Fig2}), and hence the surface charge is higher, their corresponding electrical fields, $E(R+d)=\lambda_0/[2\piup\epsilon(R+d)]$ are, of course, equal, due to the inverse dependence on the distance to the cylinder walls. However, the overcharging is higher for the $R=7.5a$ case. In the conventional overcharging, first reported for a planar electrode~\cite{jimenez04}, the overcharging would be the same for a constant surface charge density (everything else being the same), because the electrical field in this case depends only on the induced charge, as a function of the distance to the electrode, i.e., the electrical field for a plate-like electrode is $E(r)=4\piup\sigma(r)/\epsilon$, while for a cylindrical electrode this field also depends inversely on the distance to the electrode $E(r)=\lambda(r)/(2\piup\epsilon r)$. Thus, as the distance to the cylinder walls increases, the electrical field for the pore should seemingly decrease more rapidly for the $R=7.5a$ case, but we see the contrary. This implies that while the unscreened $E_0(R+d)=\lambda_0/[2\piup\epsilon (r+d)]$ in figure~\ref{Fig2}~(b) is weaker, as $r$ increases, the electrical field $E(r)=\lambda (r)/(2\piup\epsilon r)$ decreases less than that for the $R=5a$ case, due to a higher increase in its $\lambda (r)$, which would explain the higher induced charge seen in figure~\ref{Fig2}~(b). However, why is this $\lambda(r)$ higher?

\begin{figure}[!t]
\vspace{-5mm}
 \begin{minipage}{0.49\textwidth}
     \centering
\includegraphics[width=.99\textwidth]{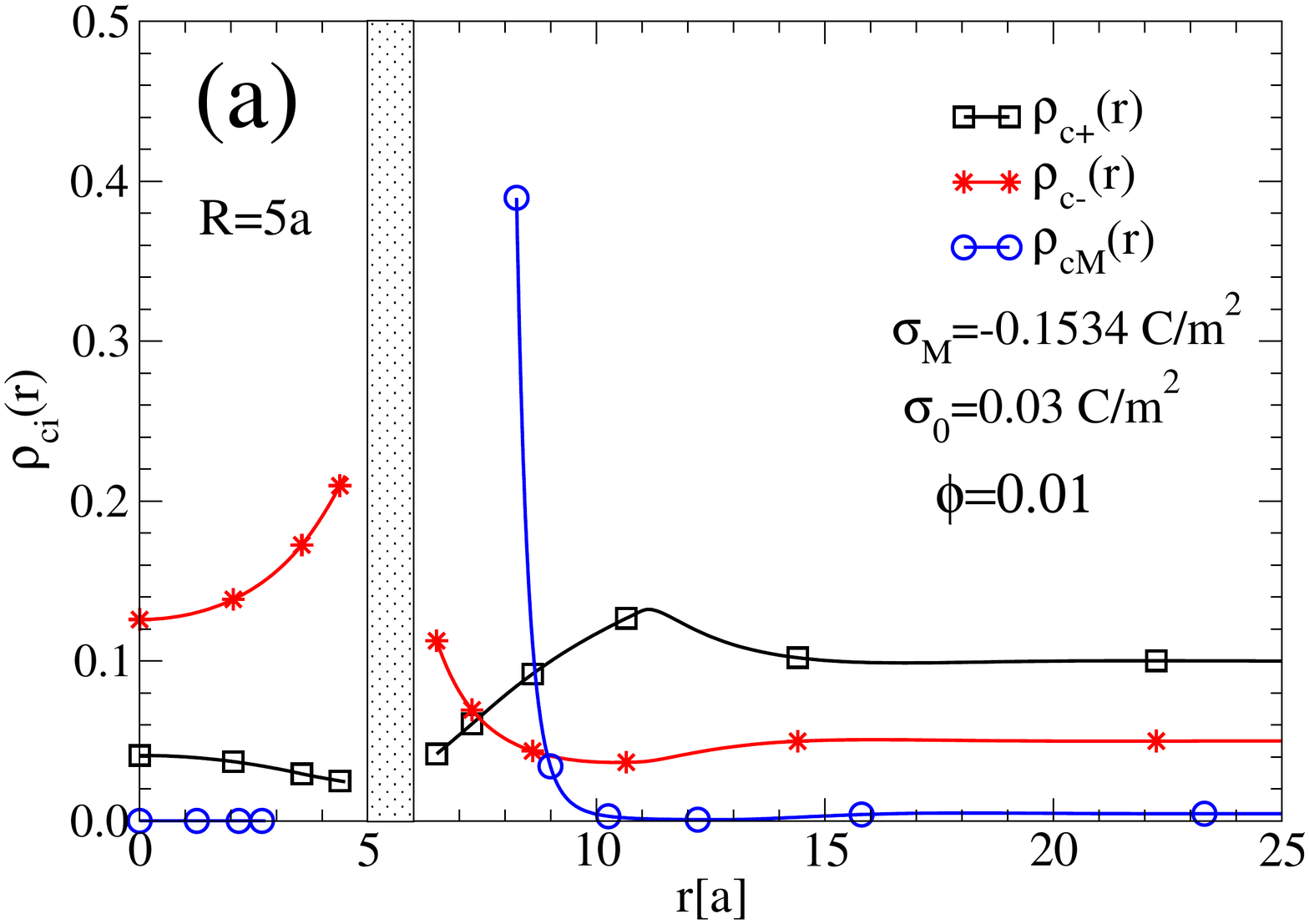}
\end{minipage}\hfill
   \begin {minipage}{0.49\textwidth}
   \centering
\includegraphics[width=.99\textwidth]{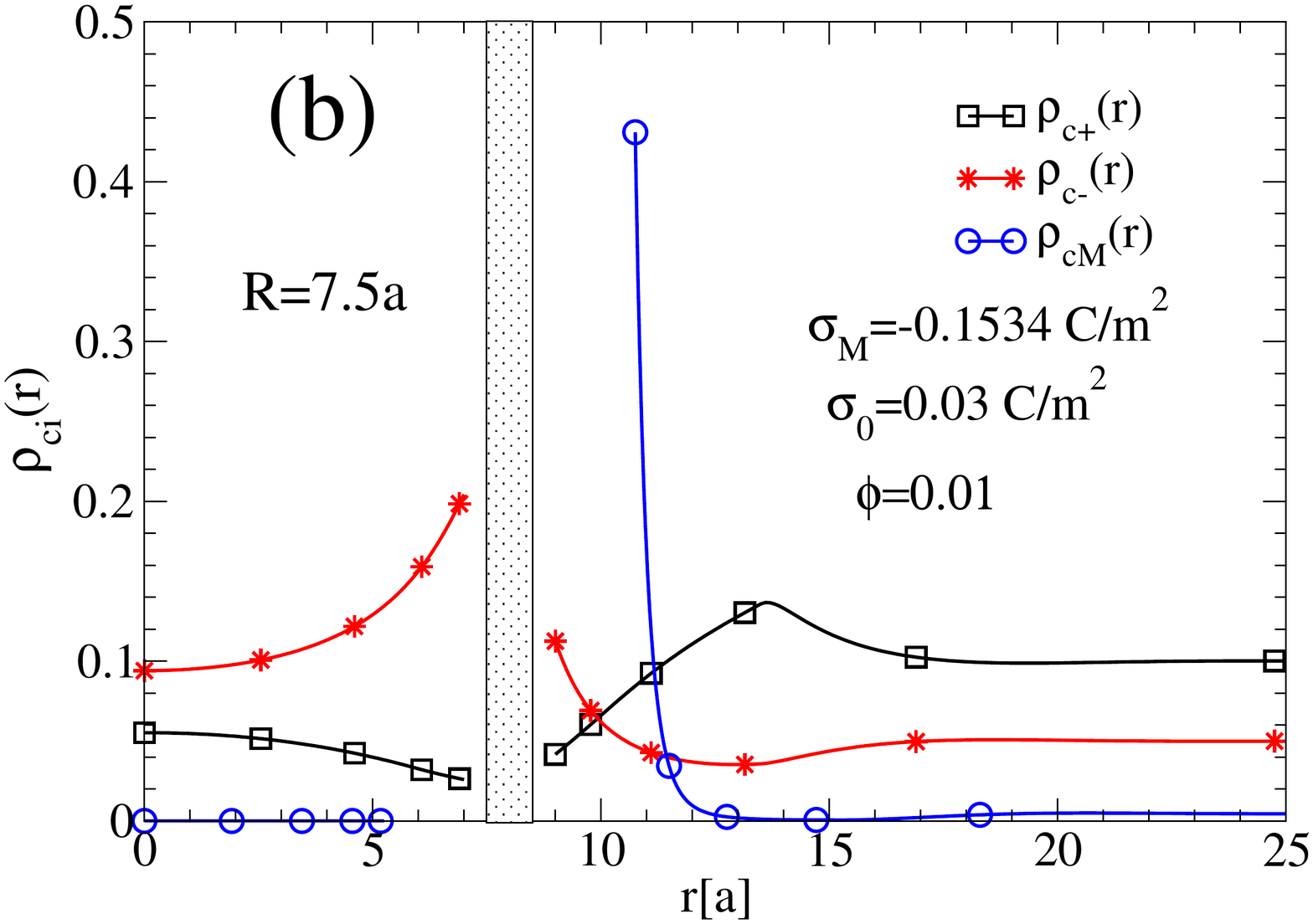}
      \end{minipage}
      \vspace{-4mm}
        \caption{\label{Fig3} (Color online) Concentration profiles in a cylindrical vesicle (pore), as a function of the distance to the central azimuthal axis, for two values of the nanopore radius. The macroions surface charge density is $\sigma=-0.1534$~C/m$^{2}$. $\rho_{c-}(r)$, $\rho_{c+}(r)$ and $\rho_{c\text{M}}(r)$ are the concentration profiles of anions, cations and macroions, in and out the cylindrical pore. Their corresponding bulk concentrations are $\rho_-^{\text{bulk}}=0.05$~M, $\rho_+^{\text{bulk}}=0.1$~M and  $\rho_{\text M}^{\text{bulk}}=0.00455$~M. All the other parameters are as in figure~\ref{Fig2}.}
        \vspace{-2mm}
\end{figure}

In figure~\ref{Fig3} we portray the corresponding concentration profiles for the induced linear charge densities in figure~\ref{Fig2}. Here, consistently with the results of the charge densities of figure~\ref{Fig2}, we see that the macroions adsorption to the vesicle is higher for the $R=7.5a$ than for the $R=5a$ case. On the other hand, we see that the adsorption of the outside cations in figure~\ref{Fig3}~(b) is a bit larger than that in figure~\ref{Fig3}~(a). Therefore, apparently the lower repulsion of the cations (associated to the macroions) for the $R=7.5a$ case produces a higher induced charge density. It should be pointed out that the distribution of cations around the macroions, suggested by the hump around the position of the closest approach of the macroions to the external cylinder wall, supports the mechanism of overcharging proposed here, i.e., the cations associated to the adsorbed macroions to the cylinder produces the overcharging. In addition, while we will not discuss here the details, we saw that there is a substantial increase, above their bulk values, of the inside anions concentration profiles. In fact, for the parameters of figure~\ref{Fig3}, the osmotic pressure, defined as $P_n=P_{\text{in}}-P_{\text{out}}$, is positive.

\begin{figure}[!t]
\vspace{-4mm}
 \begin{minipage}{0.49\textwidth}
     \centering
\includegraphics[width=.99\textwidth]{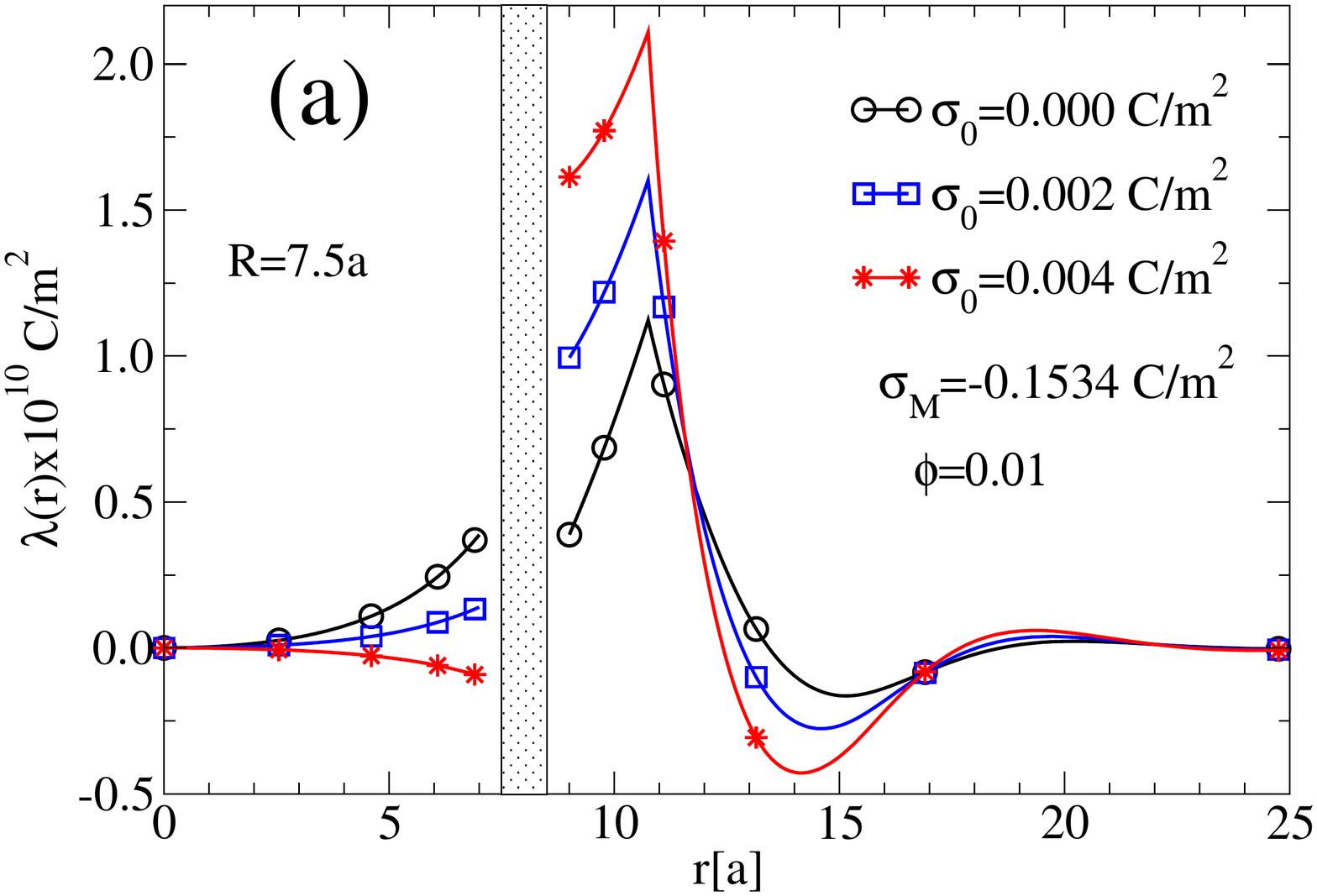}
\end{minipage}
   \begin {minipage}{0.49\textwidth}
   \centering
\includegraphics[width=.99\textwidth]{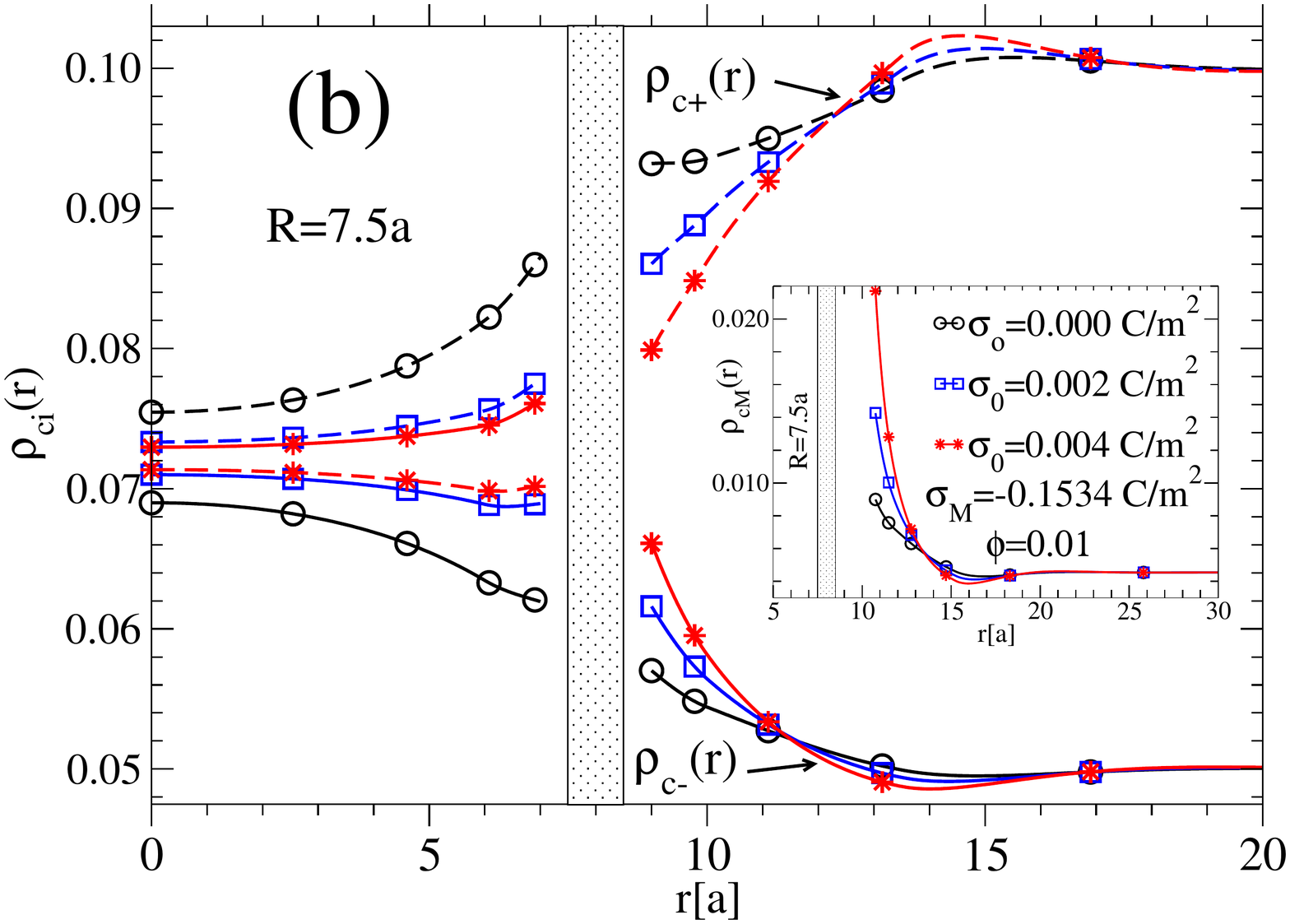}
        \end{minipage}
        \vspace{-4mm}
        \caption{\label{Fig4} (Color online) Induced charge (a) and concentration (b) profiles in a cylindrical vesicle (pore), as a function of the distance to the central azimuthal axis, for three different nanopore surface charge densities, $\sigma_0$. The macroions surface charge density is $\sigma=-0.1534$~C/m$^{2}$. $\rho_{c-}(r)$, $\rho_{c+}(r)$ and $\rho_{c\text{M}}(r)$ are the concentration profiles of anions, cations and macroions, in and out of the cylindrical pore. Their corresponding bulk concentrations are $\rho_-^{\text{bulk}}=0.05$~M, $\rho_+^{\text{bulk}}=0.1$~M and  $\rho_{\text M}^{\text{bulk}}=0.00455$~M. The radius of the nanopore is $R=7.5a$. The dashed lines are the cations concentration profiles and the solid lines are the anions concentration profiles. In the inset, we show the macroions concentration profiles. All the other parameters are as in figure~\ref{Fig2}.}
        \vspace{-2mm}
\end{figure}

In figure~\ref{Fig4}~(a) and figure~\ref{Fig4}~(b) we display the linear induced charge and concentration profiles, respectively, for $R=7.5a$, and for three low values of $\sigma_0$, i.e., $\sigma_0=0.000$~C/m$^2$, $0.002$~C/m$^2$, and $0.004$~C/m$^2$. All the other parameters are as those in figure~\ref{Fig3}~(b). For the three induced charge densities, shown in figure~\ref{Fig4}~(a), there is a significant overcharging. For the case of $\sigma_0=0.000$~C/m$^2$, we know that, as an entropy effect, the macroions are adsorbed to the pore walls, bringing with them the cations, which are responsible for the overcharging. The adsorption of the macroions on the external side of the pore also promotes the adsorption of the cations inside the pore [see figure~\ref{Fig4}~(b)], since for $\sigma_0=0.000$~C/m$^2$ there is no repulsive force, associated to the wall, on the inside cations. Thus, the symmetry breaker in this system is the entropic force which favors the adsorption of the external charged macroions to the nanopore. In the inset of figure~\ref{Fig4}~(b), we see that a higher increase of a positive $\sigma_0$ implies a greater adsorption of the negatively charged macroions to the outside wall of the nanopore, with the concomitant adsorption of cations. This mechanism is the usual explanation of the original overcharging phenomena, i.e., both the entropy force and electrical force attract oppositely charged macroions to the, either planar or cylindrical, electrode. However, notice that although there is now a positive charge on the nanopore, for the case of $\sigma_0=0.002$~C/m$^2$, the inside cations are still being attracted to the inside surface of the vesicle [see figure~\ref{Fig4}~(b)]. This is a counterintuitive result, and contrary to the results shown in figure~\ref{Fig3}  (inside the pore) for higher pore charges, where the anions are preferably adsorbed. In figure~\ref{Fig4}~(b), the adsorption of cations is present for all values of $\sigma_0\lesssim 0.004$~C/m$^2$. Such an effect is also present for the cases discussed in figures~\ref{Fig2}~(a) and \ref{Fig3}~(a) (not shown). The \textit{normal} wetting of the anions, in figure~\ref{Fig4}~(b), is recovered when $\sigma_0\geqslant0.004$~C/m$^2$. The positive induced charge, inside the pore, for $\sigma_0< 0.004$~C/m$^2$, has the effect of producing a more attractive force on the outside macroions. More interesting is the very unexpected increasing adsorption of the inside cations, as a positive $\sigma_0$ charge increases, at least up to $\sigma_0\lesssim0.004$~C/m$^2$. For $\sigma_0>0.004$~C/m$^2$, the inside anions wet the inside surface of the nanopore and, hence, the induced charge inside the pore becomes negative, then decreasing the net charge at $r=R+d$, and consequently assisting the adsorption of macroions, as the surface charge of the pore increases. The increment of macroions adsorption, in spite of an increasing induced positive charge inside the nanopore for $\sigma_0<0.004$~C/m$^2$, indicates that for this interval, the electrical field produced by the macroions adsorption penetrates the vesicle membrane, with sufficient intensity to overcome the positive charge on the membrane. Thus, showing the correlation between the fluids at both sides of the nanopore wall~\cite{Yu2,Yu3}. For $\sigma_0\geqslant 0.004$~C/m$^2$, the correlation persists, but now the net effect is to favor the macroions adsorption. We will refer to this double adsorption of cations on \textit{both} sides of the positively charged nanopore, as \textit{double overcharging}. For biological vesicles, this adsorption of inside cations on a positively charged membrane could be even more important since the dielectric constant for these systems are in general very much lower than that considered here $\epsilon =78.5$~\cite{nimeyer08}. In general, of course, the dielectric constant of the nanopore wall will modulate the intensity and the sign of the effects reported here~\cite{wang2016}.  

\section{Conclusions}

We have solved the hypernetted-chain/mean spherical approximation (HNC/MSA) integral equations for an uncharged and charged nanopore immersed into a colloidal dispersion, at finite concentration, modelled by a colloidal primitive model. We focused our calculations on the induced linear charge density and concentration profiles, inside and outside  a modelled vesicle or nanopore. We explored the dependence of these functions on the nanopore positive surface charge density ($0$~C/m$^2\leqslant\sigma_0\leqslant0.1$~C/m$^2$), and on the macroions negative charge ($z_{\text M}=-4$ and $-11$), keeping their volume fraction constant, $\phi=0.01$ ($\rho_{\text M}=0.00455$~M, $a_{\text M}=4.5a$). The salt was taken to be a 1:1, 0.05~M restricted primitive model  electrolyte, with $a=4.25$~{\AA}. We have found and report the existence of overcharging for low-to-medium-high nanopore's surface charge density, and medium-high macroions charge. For high values of $\sigma_0$, the overcharging disappears, as well as for low macroions charge density, with the exception of the very low $\sigma_0$, when it is always present. Increasing the pore radius implies a higher induced charge, inside (negative) and outside (positive). 

These phenomena seem to be counterintuitive since a higher radius implies a lower unscreened electrical field, which is inversely dependent on the distance $r$ to the nanopore. This decay is overcome by a dependence of $\lambda(r)$ at a somewhat higher power of $r$. Another interesting result is the overcharging by the cations at \textit{both sides} of a positively charged nanopore, for low values $\sigma_0$, as a result of the charge correlation of the outside fluid (adsorbed negative macroions on the outer wall of the cylinder), on the electrolyte inside the pore. This new phenomenon, to which we will refer to as \textit{double overcharging} (OC2), will be certainly enhanced for a biological membrane, where the dielectric constant is much lower than that of the water. Other factors, such as a higher colloidal volume fraction, or charge, or higher salt concentration will probably play a relevant role in this newly reported effect, by increasing the cations bulk concentration, and/or by increasing the macroions adsorption. In our model, interchanging the charge sign of the pore and the macroions will give the image results, to those reported here. We wish to finally point out the relevance of the electrolyte inside the vesicle (nanopore) as an important control mechanism of the outside induced linear charge density, i.e., as a modulator of overcharging (OC) [also referred to as the surface charge amplification (SCA)], \textit{double overcharging} (OC2), charge reversal (CR) and charge inversion (CI).

\section*{Acknowledgements}

The support of CONACYT (Grant No.~169125) is gratefully acknowledged. E. G.-T. thanks the assistance of the computer technicians at the IF-UASLP.

\vspace{-4mm}
\ukrainianpart
\title{Модельна заряджена циліндрична нанопора в  колоїдній дисперсії: зарядоінваріантність, явища надмірного зарядження і подвійного
надмірного зарядження}

\author{E. Ґонзалез-Товар\refaddr{label1}, М. Лозада-Кассу\refaddr{label2}}

\addresses{
\addr{label1} Інститут фізики, Автономний університет Сан-Луїс-Потосі, Сан-Луїс-Потосі, Мексика
\addr{label2} Інститут відновлювальних джерел енергії, Нацiональний автономний унiверситет м. Мехiко,\\  Мехіко, Мексика
}

\makeukrtitle

\begin{abstract}

Використовуючи гіпернет-ланцюгове/середньо-сферичне наближення інтегральних рівнянь, ми вивчаємо подвійний електричний шар всередині
і зовні модельної зарядженої циліндричної бульбашки (нанопори), зануреної в примітивну модель розчину макроіонів таким чином, 
що макроіони присутні лише зовні нанопори, тобто стінка бульбашки є проникною лише для зовнішніх макроіонів.    
Ми обчислюємо профілі іонної густини і локальної лінійної зарядової густини всередині і зовні бульбашки і знаходимо, що 
кореляція між внутрішнім і зовнішнім розподілами спричинює явище надмірного зарядження  (також відоме як збільшення поверхневого заряду)
і/або  зарядоінваріантність. Це вперше, коли явище надмірного зарядження прогнозується в електричному подвійному шарі циліндричної 
геометрії.
Ми також повідомляємо про нове явище  подвійного  надмірного зарядження.
Представлені результати можуть мати значення для пов'язаних систем у фізичній хімії, для збереження енергії, в біології, а саме, для 
нанофільтрів, конденсаторів і клітинних мембран.

\keywords бульбашки, надмірне зарядження, подвійне надмірне зарядження, енергія, батареї

\end{abstract}

\end{document}